\documentclass{article}
\usepackage[utf8]{inputenc}
\usepackage[T1]{fontenc} 
\usepackage{arxiv}
\usepackage{amsmath}
\usepackage{braket}
\usepackage{amsfonts}
\usepackage{amssymb}
\usepackage{graphicx}
\usepackage{amsthm}
\usepackage{listings}
\usepackage{float}
\usepackage{caption}
\usepackage{subcaption}
\usepackage{color}
\usepackage[colorlinks=true,allcolors=blue]{hyperref}
\usepackage{chngcntr}
\usepackage{cite}
\usepackage{authblk}
\usepackage{tikz}
\usepackage{booktabs} 
\DeclareFontFamily{OMS}{oasy}{\skewchar\font48 }
\DeclareFontShape{OMS}{oasy}{m}{n}{%
	<-5.5> oasy5     <5.5-6.5> oasy6
	<6.5-7.5> oasy7     <7.5-8.5> oasy8
	<8.5-9.5> oasy9     <9.5->  oasy10
}{}
\DeclareFontShape{OMS}{oasy}{b}{n}{%
	<-6> oabsy5
	<6-8> oabsy7
	<8->  oabsy10
}{}
\DeclareSymbolFont{oasy}{OMS}{oasy}{m}{n}
\SetSymbolFont{oasy}{bold}{OMS}{oasy}{b}{n}

\DeclareMathSymbol{\smallleftarrow}     {\mathrel}{oasy}{"20}
\DeclareMathSymbol{\smallrightarrow}    {\mathrel}{oasy}{"21}
\DeclareMathSymbol{\smallleftrightarrow}{\mathrel}{oasy}{"24}

\definecolor{greennew}{rgb}{0.1, 0.3, 0.5}
\definecolor{rednew}{rgb}{0.8, 0, 0}
\definecolor{bluenew}{rgb}{0, 0, 0.8}
\definecolor{newgreen}{rgb}{0.07, 0.53, 0.37}

\usepackage{mathptmx}
\renewcommand{\vec}[1]{\ensuremath{\boldsymbol{#1}}}

\newcommand{\sgn}{\,\mbox{\rm sgn}}

\counterwithout{figure}{section}

\title{Extended transfer matrix method for electron transmission in anisotropic 2D materials: Interplay of strain and (a)periodicity of potentials}

\author{E D\'iaz-Bautista\textsuperscript{1}, Y Betancur-Ocampo\textsuperscript{2}, A Raya\textsuperscript{3,4}\\
\textsuperscript{1} Instituto Polit\'ecnico Nacional, UPIIH, Ciudad del Conocimiento y la Cultura, 42162 Hidalgo, Mexico \\
\textsuperscript{2} Instituto de Física, Universidad Nacional Autónoma de México, Ciudad de México, México \\
\textsuperscript{3} Instituto de F\'{\i}sica y Matem\'aticas, Universidad Michoacana de San Nicol\'as de Hidalgo, Edificio C-3, Ciudad Universitaria. Francisco J. M\'ujica s/n. Col. Fel\'{\i}citas del R\'{\i}o. 58040 Morelia, Michoac\'an, M\'exico\\
\textsuperscript{4} Centro de Ciencias Exactas, Universidad del Bío-Bío. Avda. Andrés Bello 720, Casilla 447, 3800708, Chillán, Chile \\
e-mail: ediazba@ipn.mx, ybetancur@fisica.unam.mx, alfredo.raya@umich.mx}
\date{ }

\begin{document}

\maketitle
\begin{abstract}
We extend the conventional transfer matrix method to include anisotropic features for electron transmission in two-dimensional materials, such as breaking reflection law in pseudo-spin phases and wave vectors. This method allows to study transmission properties of anisotropic and stratified electrostatic potential media from a wide range of tunable parameters, which include strain tensor and gating. We apply the extended matrix method to obtain the electron transmission, conductance, and Fano factor for the interplay of an uniaxially strained graphene sheet with external one-dimensional aperiodic potentials. Our results suggest the possibility of visualizing this interplay from conductance measurements.
\end{abstract}

Graphene~\cite{Novoselov666,Novoselovgraphene,Zhang2005ExperimentalOO,GeimRise},  the wonder material, possesses an unusual gapless band structure which makes it an ideal candidate for technological applications. Since its discovery, several band engineering techniques have been developed to modify its electronic properties, many of which have been extended to the family of all 2D materials. On one hand, straintronics~\cite{Pereira09} or origami electronics~\cite{TOMANEK200286} emerged as the field devoted to manipulate the electronic properties of graphene by mechanical deformations~\cite{Naumis_2017,diaz20,Pereira09,guinea09,yonatan18,yonatan21,contreras19,lima16,Chauwin2022,DiazBautista2020,Antonova2021,Concha2018,Mannai2020,Le2020,Phan2021}. Strain can also appear by placing the graphene sample on a substrate with a misaligned lattice, giving rise to superlattices with associated Moiré patterns~\cite{Artaud16}. Strain engineering is naturally extended to other materials, including those which are naturally anisotropic~\cite{Naumis_2017,Katayama2009ElectronicPC,Kajita14}.

In general, strain configurations induce  pseudomagnetic fields. A triaxial strain, for instance, is capable of producing a uniform magnetic field that induces Landau level quantization with pseudomagnetic fields of up to 50~Tesla~\cite{guinea09}. At the same time, such strain induces gap opening in the superlattice and pseudo-Landau level quantization. Both periodically and quasiperiodically structures offer a plethora of possibilities in 2D materials for technological applications \cite{Tsu73,Tsu74,Beltram88,wang10,lejarreta13,barbier08,barbier09,agrawal13,XU2015188,ramezani10,Park2008,BEZERRA2020,Naumis2019,baake_grimm_2017,cjanot2012,GARCIACERVANTES2015,FELIX2020,garciacervantes2017,Guzman2018,MolinaValdovinos2022,Miniya2022,Carrillo2021,GARCIACERVANTES201599,doi:10.1063/1.4729133,doi:10.1063/1.4772209,doi:10.1063/1.4788676,Xu_2013,doi:10.1063/1.4826643,Korol2018}. In this regard, an aperiodic modulation of stratified electrostatic potential media can be considered to control the transmission properties of devices ~\cite{GARCIACERVANTES2015}. In graphene, aperiodic order arrays in the form of Cantor~\cite{garciacervantes2017}, Fibonacci~\cite{FELIX2020}, and Thue-Morse~\cite{doi:10.1063/1.4729133,doi:10.1063/1.4772209,doi:10.1063/1.4788676,Xu_2013,doi:10.1063/1.4827380,doi:10.1063/1.4826643,doi:10.1063/1.4868529,doi:10.1063/1.4757591,BRIONESTORRES201498,CARRERAESCOBEDO2014248} as well as periodic arrays~\cite{GARCIACERVANTES2022414052,Chen_2013,Zhang_2014,ZHANG20141413} have been intensively explored~\cite{GARCIACERVANTES201599} in connection with the transmission properties of  electrons along 2D materials. Most of the transmission properties are dependent on incident angle and minibands and minigaps emerge, where electron filtering is feasible. Moreover, an aperiodic order reduces the number of resonant peaks in the transmission probability, making it possible to establish pass and/or stop bands.

In this letter, we reformulate the matrix transfer method to introduce properly the wave vector and pseudo-spin angles for situations where the dispersion relation is anisotropic. We explore the issue of resonant tunneling in a graphene superlattice for homogeneous strain, avoiding the generation of any pseudomagnetic field whatsoever. For this purpose, we use a two level band model for anisotropic two-dimensional materials, which is a natural generalization of previous approaches for isotropic systems \cite{lima16,walker94,GARCIACERVANTES2015,BRIONESTORRES201498}. We develop a transfer matrix setup (see \cite{walker94} for a pedagogical introduction to the subject) and compute the transmission coefficient in terms of the controlling parameters of the model, namely, the modulated Fermi velocity itself, and the height of the barriers in stratified electrostatic potential media. We further introduce a tunable parameter that permits transit between periodic and aperiodic orders of the strata. Fibonacci sequence is considered to illustrate the impact of this parameter in the transmission under consideration.  

\section{The model}\label{sec2}
We start with a general model of two level system with symmetric conduction and valence bands, which can be based on tight-binding (TB) approach to nearest neighbors consisting of a $2\times 2$ Hamiltonian given by
\begin{equation}\label{2levels}
H_{\rm TB}  = \left(\begin{array}{c c}
V & g^*(\vec{k}) \\
g(\vec{k}) & V
\end{array}\right)=V\sigma_0+\sigma_x g_x(\vec{k}) + \sigma_yg_y(\vec{k}),
\end{equation}
\noindent where the function $g(\vec{k}) = g_x(\vec{k}) + i g_y(\vec{k})$ and $\vec{k} = (k_x, k_y)$ is the wave vector. We consider a constant electrostatic potential $V$, whose effect is to arise the dispersion relation in the energy scale. Here, $\sigma_x$ and $\sigma_y$ are the Pauli matrices and $\sigma_0$ is the $2\times2$ identity matrix. Energy bands and eigenfunctions are 
\begin{equation}\label{bands}
E = V + s|g(\vec{k})|,
\end{equation}
\noindent and
\begin{equation}\label{wave}
\vec{w}({\vec{r},\vec{k}})=\frac{1}{\sqrt{2}}\left(\begin{array}{c}
1 \\
s \exp\left(i\phi(\vec{k})\right)
\end{array}\right)\exp\left(i\vec{k}\cdot{\bf r}\right),
\end{equation}
\noindent respectively, where $s = \sgn(E - V)$ is the band index and the pseudo-spin angle is given by 
\begin{equation}\label{pseudospin}
    \phi(\vec{k}) = \arctan\left(\frac{g_y(\vec{k})}{g_x(\vec{k})}\right).
\end{equation}

\begin{figure}
	\centering
	\includegraphics[width=0.5\textwidth]{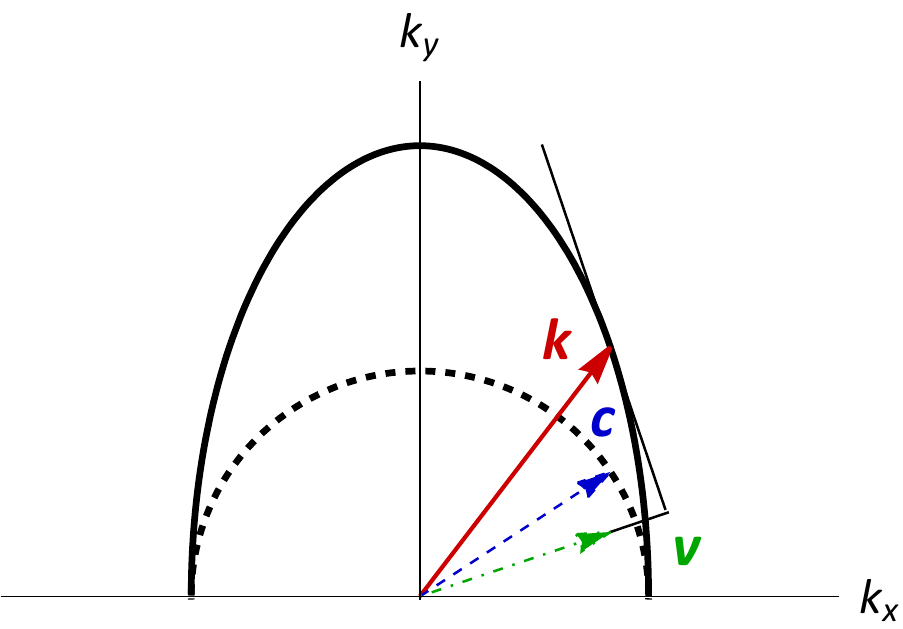}
	\caption{Schematic representation of the wave vector $\vec{k}$ (red arrow), the generalized pseudo-spin $\vec{c}$ (dashed blue arrow), and the group velocity $\vec{v}$ (dash-dotted green arrow) of anisotropic dispersion relations in 2D materials.}
	\label{fig:fig0}
\end{figure}

The function $g(\vec{k})$ changes on the two-dimensional material. It is important to note that this pseudo-spin angle is different to the wave vector and group velocity direction (see Fig. \ref{fig:fig0}), which are given, respectively, by
\begin{equation}
    \gamma(\vec{k}) = \arctan\left(\frac{k_y}{k_x}\right),
\end{equation}
and
\begin{equation}\label{incangle}
    \theta(\vec{k}) = \arctan\left(\frac{\partial_{k_y}|g(\vec{k})|}{\partial_{k_x}|g(\vec{k})|}\right).
\end{equation}

In the particular case of graphene, with $g(\vec{k}) = \hbar v_{\rm F}(k_x + ik_y)$, we can obtain $\phi(\vec{k}) = \gamma(\vec{k}) = \theta(\vec{k})$. However, such an equality is not valid generally for anisotropic materials.

Hamiltonian in Eq.~\eqref{2levels} embodies two famous cases of anisotropic materials: The first one corresponds to uniaxially strained graphene, where the function $g(\vec{k})$ from TB to nearest neighbors has the form
\begin{equation}\label{gstr}
    g(\vec{k}) = \sum^3_{j=1}t_j \textrm{e}^{i\vec{k}\cdot\vec{\delta}_j},
\end{equation}
\noindent where hopping parameters $t_j$ up to nearest neighbors follow the exponential decay rule~\cite{Pereira09}
\begin{equation}
    t_j = t\exp\left(\frac{\delta_j}{a}-1\right),
\end{equation}
\noindent with $t = 2.71$ eV and $a = 0.142$ nm. The deformed bond lengths $\delta_j$ and nearest neighbor positions $\vec{\delta}_j$ are modulated in terms of strain tensor parameters, as in Ref.~\cite{Pereira09}.

The electronic band structure of uniaxially strained graphene has two pseudo-spin valleys with elliptical Dirac cones near the corner of the first Brillouin zone in low energy regime. This semimetallic phase remains if the parameters $t_j$ satisfy the triangular inequality $||t_1|-|t_3||\leq t_2 \leq |t_1 + t_3|$. The second case corresponds to phosphorene whose electronic band structure shows an insulating phase at the center of the first Brillouin zone. Pseudo-spin valleys in phosphorene disappear due to the fact that the hopping parameters $t_1 = t_3 = -1.2$ and $t_2 = 3.4$ do not satisfy the triangular inequality of hexagonal lattices \cite{Pereira09}.

\begin{figure}
	\centering
	\includegraphics[width=\textwidth]{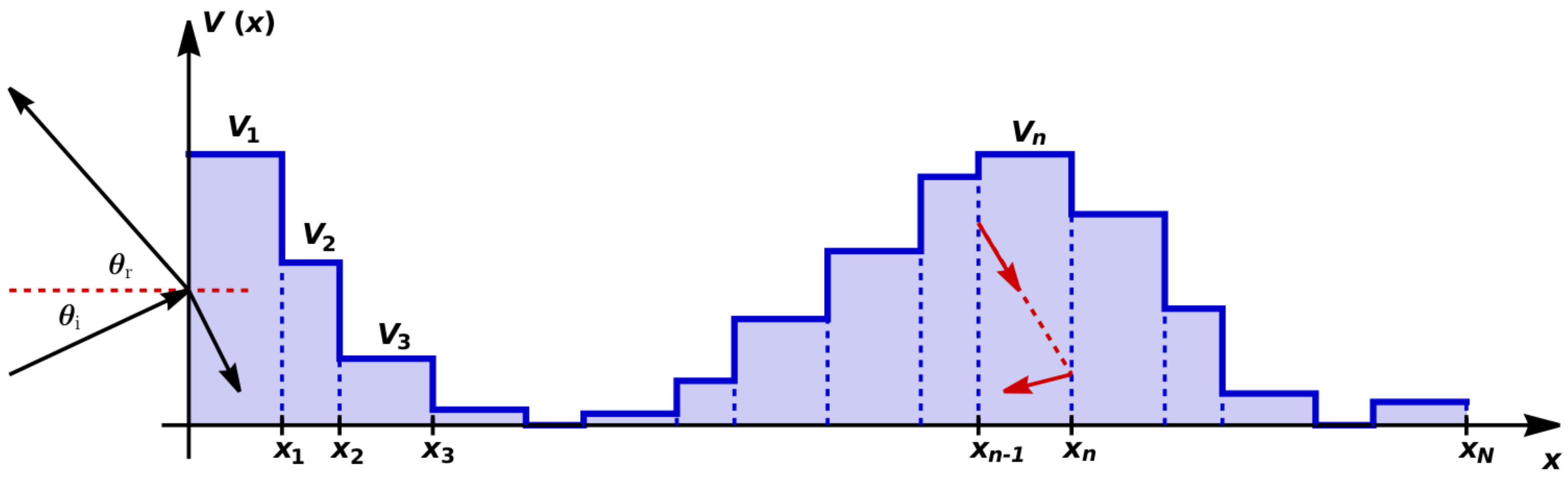}
	\caption{Schematic representation of the finite periodic potential barriers along the $x$ axis. The width of each barrier is $D$ and the separation between two consecutive barriers is $L$.}
	\label{fig:fig1}
\end{figure}
 
\section{Extended transfer matrix method}
We study the electron scattering of anisotropic materials considering stratified electrostatic potential media, which consists of adjacent potential barriers of width $x_n - x_{n-1}$ and height $V_n$, being $n$ the label of the adjacent barrier, as shown in Fig. \ref{fig:fig1}. The modification to the conventional matrix method starts proposing the wave function ansatz inside the $n$-th barrier as
\begin{equation}
    \vec{\Psi}^n(\vec{r}) = t_{n}\vec{w}(\vec{r},\vec{k}^t_n) + r_{n}\vec{w}(\vec{r},\vec{k}^r_n),
\end{equation}
\noindent where $t_{n}$ and $r_{n}$ are the amplitudes of the transmitted and reflected wave, respectively. In the case $n = 0$ for the first region, $t_0 = 1$ that corresponds to the incident wave, with an incident wave vector $\vec{k}^t_0$, and $r_0 = r$ for the reflected part and wave vector $\vec{k}^r_0$. With $n = N+1$ in the last region, $r_{N+1} = 0$ because there is no reflection and $t_{N+1} = t$ for the outcoming wave. The spinors $\vec{w}(\vec{r},\vec{k}^{(t/r)}_n)$ have the same expression as in Eq.~\eqref{wave} setting $\vec{k}^{(t/r)}_{n}$. These wave vectors can be parametrized in terms of the scattering angle using Eqs. \eqref{bands} and \eqref{incangle} for the particular material studied.

Applying the continuity condition at $x = x_{n}$ that separates the regions with wave functions $\vec{\Psi}^n(\vec{r})$ and $\vec{\Psi}^{n+1}(\vec{r})$, we have
\begin{equation}
    \left(\begin{array}{c}
          t_{n+1}\\
          r_{n+1}
    \end{array}\right) = M^{-1}_{n+1 n} M_{n n}\left(\begin{array}{c}
          t_{n}\\
          r_{n}
    \end{array}\right),
\end{equation}
\noindent where the matrices $M_{m l}$ are
\begin{equation}
    M_{m l} = \frac{1}{\sqrt{2}}\left(\begin{array}{cc}
      \textrm{e}^{ik^t_{x,m} x_l}  & \textrm{e}^{ik^r_{x,m} x_l} \\
        s_m \textrm{e}^{i(\phi(\vec{k}^t_m)+k^t_{x,m} x_l)} & s_m \textrm{e}^{i(\phi(\vec{k}^r_m)+k^r_{x,m} x_l)}
    \end{array}\right),
\end{equation}
\noindent which take into account the conservation of $k_y$ parallel to the interfaces at $x = x_n$. We find a $2\times2$ linear  system of equations for the amplitudes $r$ and $t$ defining
\begin{equation}
\Lambda = \prod^N_{n=1} M_{n n-1}M^{-1}_{n n},
\end{equation}
\noindent where
\begin{equation}
    M_{n n-1}M^{-1}_{n n} =\frac{1}{\textrm{e}^{i\phi(\vec{k}^r_n)}-\textrm{e}^{i\phi(\vec{k}^t_n)}} \left(\begin{array}{cc}
      s_n[\textrm{e}^{i\phi(\vec{k}^r_n)-\xi^t_{n n-1}}-\textrm{e}^{i\phi(\vec{k}^t_n)-\xi^r_{n n-1}}]  & \textrm{e}^{-i\xi^r_{n n-1}}-\textrm{e}^{-i\xi^t_{n n-1}}\\
        \textrm{e}^{i(\phi(\vec{k}^t_n)-\phi(\vec{k}^r_n))}(\textrm{e}^{-i\xi^t_{n n-1}}-\textrm{e}^{-i\xi^r_{n n-1}}) & s_n[\textrm{e}^{i\phi(\vec{k}^r_n)-\xi^r_{n n-1}}-\textrm{e}^{i\phi(\vec{k}^t_n)-\xi^t_{n n-1}}]
    \end{array}\right)\;,
\end{equation}
\noindent being $\xi^{(r/t)}_{n n-1} = k^{(r/t)}_{x,n}(x_n - x_{n-1})$ the phase shift for the outgoing wave in the $n$-th barrier. The matrix $\Lambda$ contains all the information of electron scattering inside the stratified electrostatic potential media. Therefore, the linear system is similar to $pn$ junctions because the effect of $\Lambda$ is to replace the stratified media by a single interface at $x = 0$, as written below
\begin{equation}
    \Lambda\left(\begin{array}{c}
         1\\
         s_{N+1}\textrm{e}^{i\phi(\vec{k}^t_{N+1})}
    \end{array}\right)\textrm{e}^{ik^t_{x,N+1}x_N} t=\left(\begin{array}{c}
         1\\
         s_{0}\textrm{e}^{i\phi(\vec{k}^t_{0})}
    \end{array}\right) + r \left(\begin{array}{c}
         1\\
         s_{0}\textrm{e}^{i\phi(\vec{k}^r_{0})}
    \end{array}\right).
\end{equation}
\noindent Solving to find the squared modulus of transmission amplitude $t$, we have
\begin{equation}\label{tsquar}
    |t|^2 = \frac{4\sin^2\left[\frac{1}{2}(\phi(\vec{k}^t_{0}) - \phi(\vec{k}^r_{0})\right])}{(\rho_y-s_0\rho_x\textrm{e}^{i\phi(\vec{k}^r_0)})^2},
\end{equation}
\noindent where
\begin{equation}
    \left(\begin{array}{c}
         \rho_x\\
         \rho_y 
    \end{array}\right) = \Lambda \left(\begin{array}{c}
        1\\
        s_{N+1}\textrm{e}^{i\phi(\vec{k}^t_{N+1})}
    \end{array}\right).
\end{equation}
\noindent The transmission coefficient is obtained from the conservation of the current probability density $J_x$,
\begin{equation}\label{transm}
    T = \frac{\cos\phi(\vec{k}^t_{N+1})}{\cos\phi(\vec{k}^t_{0})}|t|^2.
\end{equation}
\noindent It is worth to mention that the procedure described above generalizes known results for graphene in $pn$ junctions, transistors, and one-dimensional superlattices~\cite{Allain2011,Katsnelson2006,Park2008}. The conventional expression for the transmission in isotropic systems as graphene occurs when $\phi(\vec{k}^t_0) = \pi - \phi(\vec{k}^r_0)$ in Eq. \eqref{tsquar}. It is not unusual that this reflection law for pseudo-spin phases is used directly in anisotropic materials, which is wrong, generally speaking. Uniaxially strained graphene and phosphorene present negative and anomalous reflection, as well as partial positive refraction in asymmetric Veselago lenses \cite{BetancurOcampo2019,yonatan18,yonatan21}. An accurate calculation of the transmission probability in Eq.~\eqref{transm} requires the derivation of the pseudo-spin phase $\phi(\vec{k}^r_0)$ from Eq. \eqref{pseudospin}. Therefore, the extension of the transfer matrix method consists in considering the adequate pseudo-spin angles in the scattering process of anisotropic materials. As example, we study the interplay of uniaxial strain and one-dimensional aperiodic potentials in graphene.

\section{Interplay of strain and quasi crystallinity in graphene}
The expansion of energy bands in Eq. \eqref{bands} around a Dirac point is
\begin{equation}\label{dr}
     E= V_n +s\hbar\sqrt{k_{x}^2v_{x}^2+2k_{x}k_{y}v_{x}^2\tan(\theta_{\rm KT})+k_{y}^2v_{y}^2},
 \end{equation}
 \noindent where $v_x$ and $v_y$ are strain-dependent Fermi velocities. The quantity $\theta_{\rm KT}$ is the direction of the anomalous Klein tunneling, which is proportional to $\epsilon \sin2\zeta$ being $\epsilon$ the tensile parameter and $\zeta$ the tension angle \cite{yonatan21}. Equation \eqref{dr} corresponds to the dispersion relation of an electron inside the $n$-th barrier. With the conservation of $k_y$, we can get exact expressions for $k^{(t/r)}_{n,x}$ using Eq.~\eqref{dr}, namely,
 \begin{equation}
     k^{(t/r)}_{n,x} = (+/-)s_n\sqrt{\frac{(E - V_n)^2}{\hbar^2v^2_x} - k^2_y\left(\frac{v^2_y}{v^2_x} - \tan^2\theta_{\rm KT}\right)}  -k_y\tan\theta_{\rm KT},
 \end{equation}
 \noindent which are necessary as input in the extended matrix transfer method and useful to calculate the pseudo-spin angle $\phi(\vec{k})$ from Eq. \eqref{pseudospin}.
 The parametrization of $k_y$ in terms of incidence angle $\theta$ can be obtained using the dispersion relation in Eq. \eqref{dr} and the group velocity direction in Eq. \eqref{incangle}, yielding
 \begin{equation}
     k_y = \frac{|E|(\tan\theta - \tan\theta_{\rm KT})}{v_x\hbar\left(\frac{v^2_y}{v^2_x}-\tan^2\theta_{\rm KT}\right)\sqrt{1+\frac{(\tan\theta - \tan\theta_{\rm KT})^2}{\frac{v^2_y}{v^2_x}-\tan^2\theta_{\rm KT}}}},
 \end{equation}
\noindent where $v_x$ and $v_y$, as well as the Klein tunneling angle $\theta_\textrm{KT}$, are expressed in terms of the strain tensor in Ref.~\cite{yonatan18}.

To complete the interplay of strain and one-dimensional stratified electrostatic potential media, we adopt the method discussed in~\cite{Naumis2019} to generate unit cells of superlattices that follow an aperiodic configuration, such as quasicrystals or a Fibonacci sequence~\cite{baake_grimm_2017,cjanot2012}. To generate the positions $x_n$ for the interfaces of the aperiodic electrostatic potential, we use the relation
\begin{equation}\label{sequence}
    x_{n}=r(n\cos\alpha+\lfloor n\tan\alpha\rfloor \sin\alpha),
\end{equation}
where $r$ is a scaling length, which is set at the value of 5 nm for our proposes. The operation $\lfloor\cdot \rfloor$ represents the integer part of a number. The sequence in Eq. \eqref{sequence} is quasi-periodic when $\tan\alpha$ is irrational; the famous Fibonacci chain is obtained when $\tan\alpha=(\sqrt{5}-1)/2=\varphi^{-1}$, being $\varphi$ the golden ratio.

\begin{figure}
	\centering
	\includegraphics[width=\linewidth]{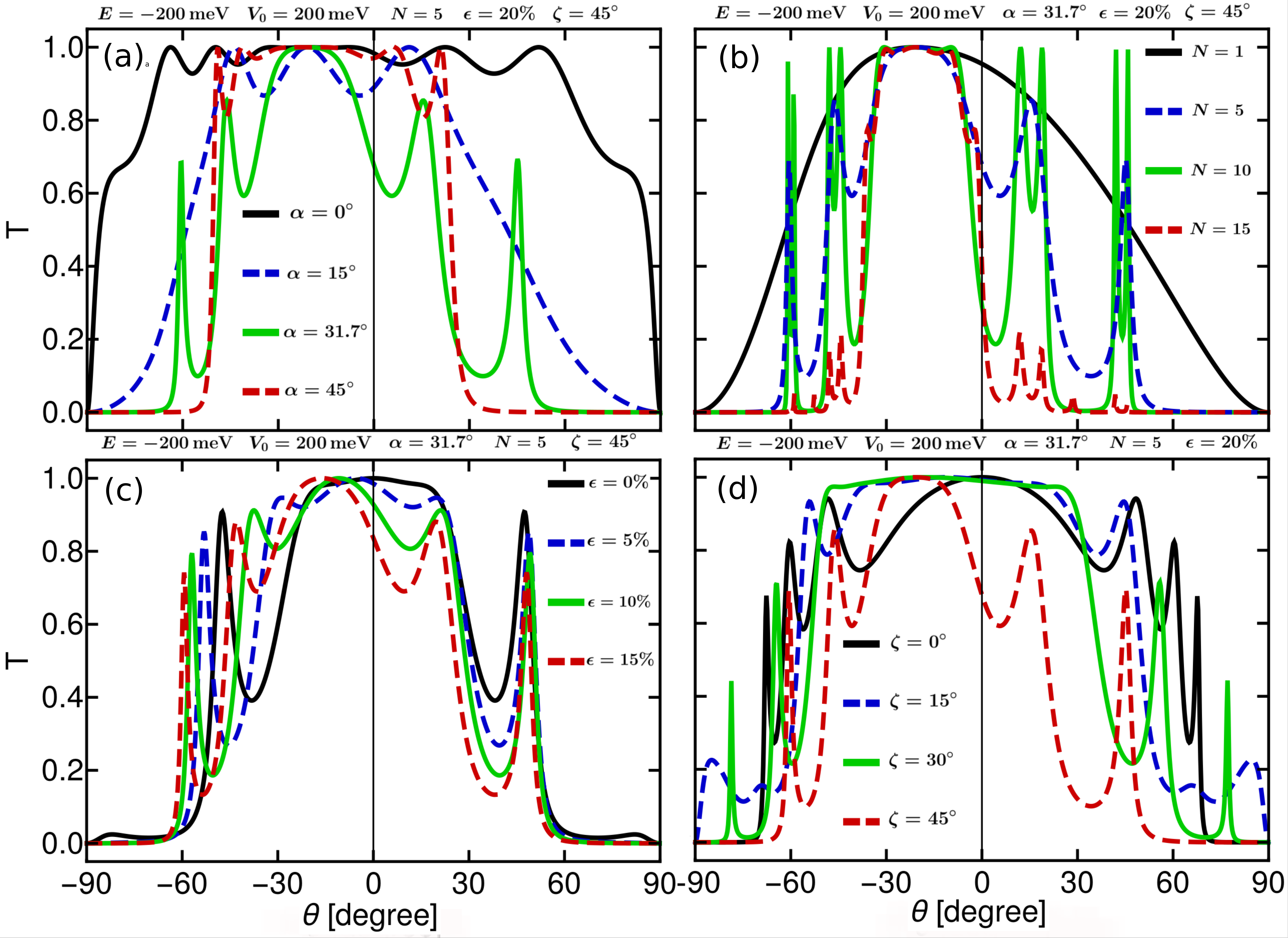}
	\caption{Electron transmission $T$ as a function of incidence angle $\theta$ for an incidence energy $E=-200$ meV, potential height $V_{0}=200$ meV, and different values of the parameter $\alpha$ (a), number $N$ of barriers (b), strain parameter $\epsilon$ (c) and tension angle $\zeta$ (d).}
	\label{fig:fig2}
\end{figure}

\begin{figure}
	\centering
	\begin{minipage}[b]{\textwidth}
		\includegraphics[width=\textwidth]{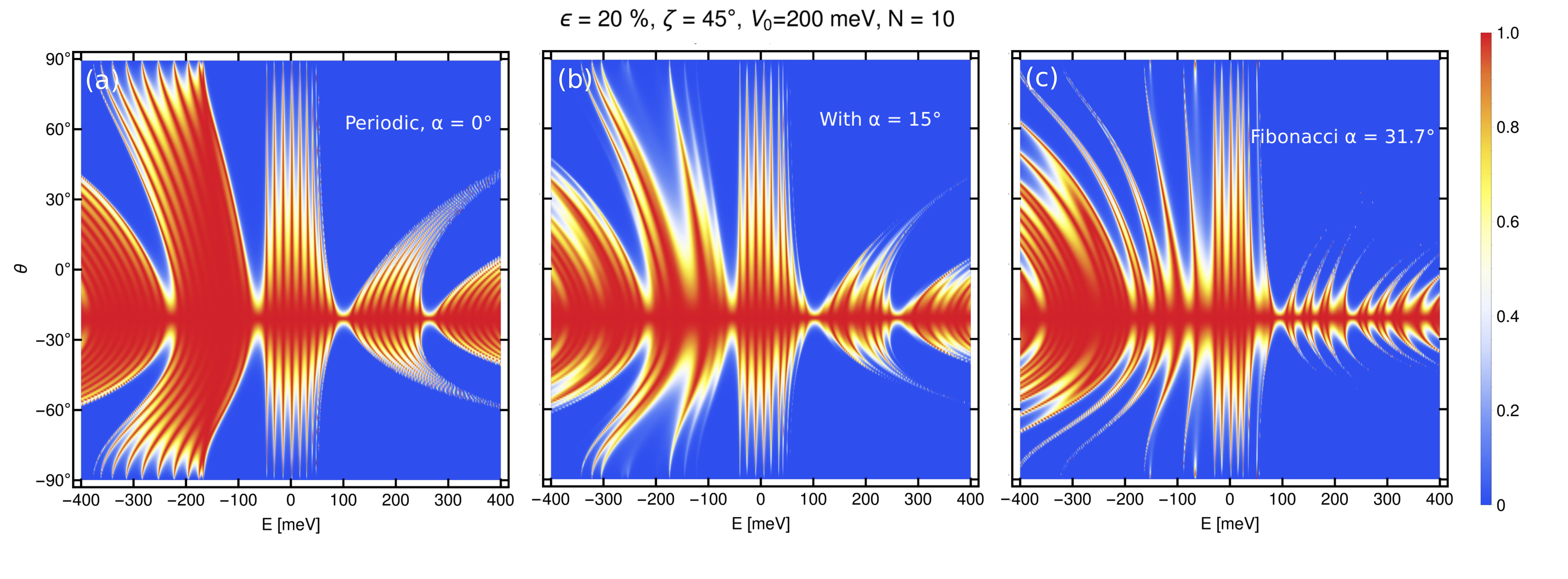}
		\label{fig:fig3a}
	\end{minipage}
	\hspace{1cm}
	\begin{minipage}[b]{\textwidth}
		\includegraphics[width=\textwidth]{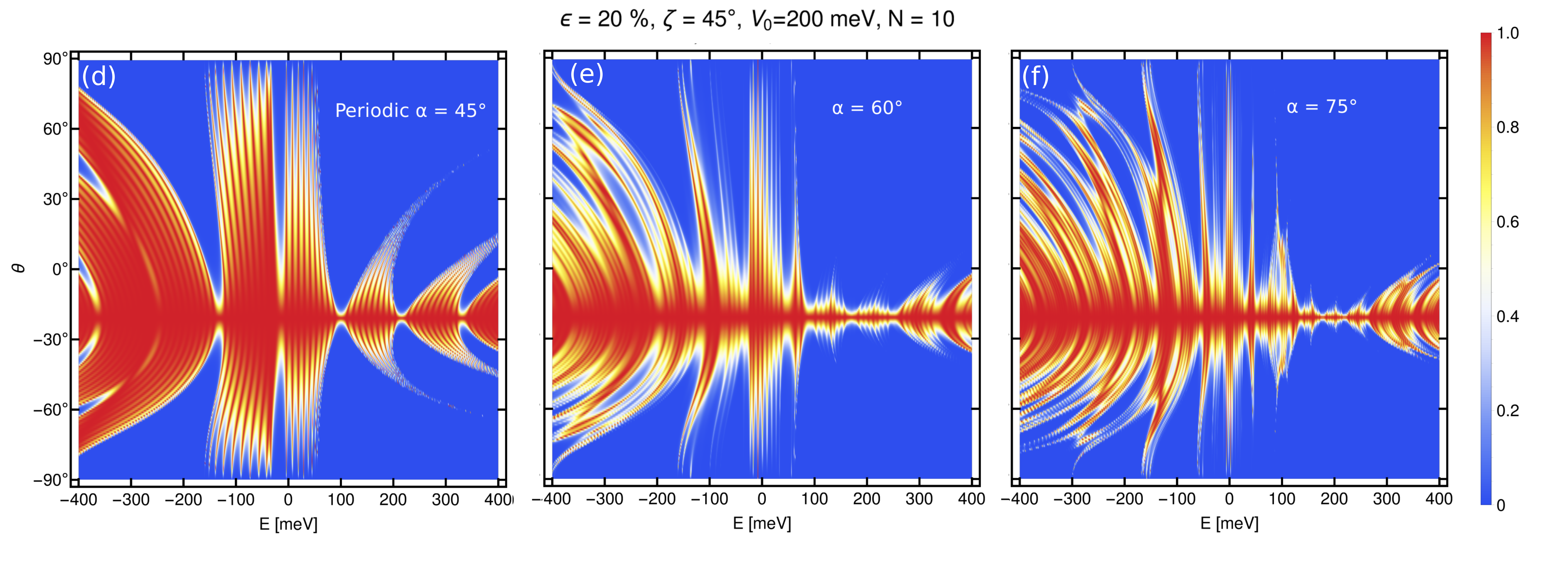}
		\label{fig:fig3b}
	\end{minipage}
	\caption{\label{fig:fig3}(a)-(f) Electron transmission $T$ as function of the Fermi level$E$ and incidence angle $\theta$ for 10 barriers and different values of the parameter $\alpha$. We set the values of the barrier height as $V_{0}=200$ meV, tensile strain $\epsilon=20\%$ and tension angle $\zeta=45^\circ$.}
\end{figure}

\begin{figure}
	\centering
	\includegraphics[width=\textwidth]{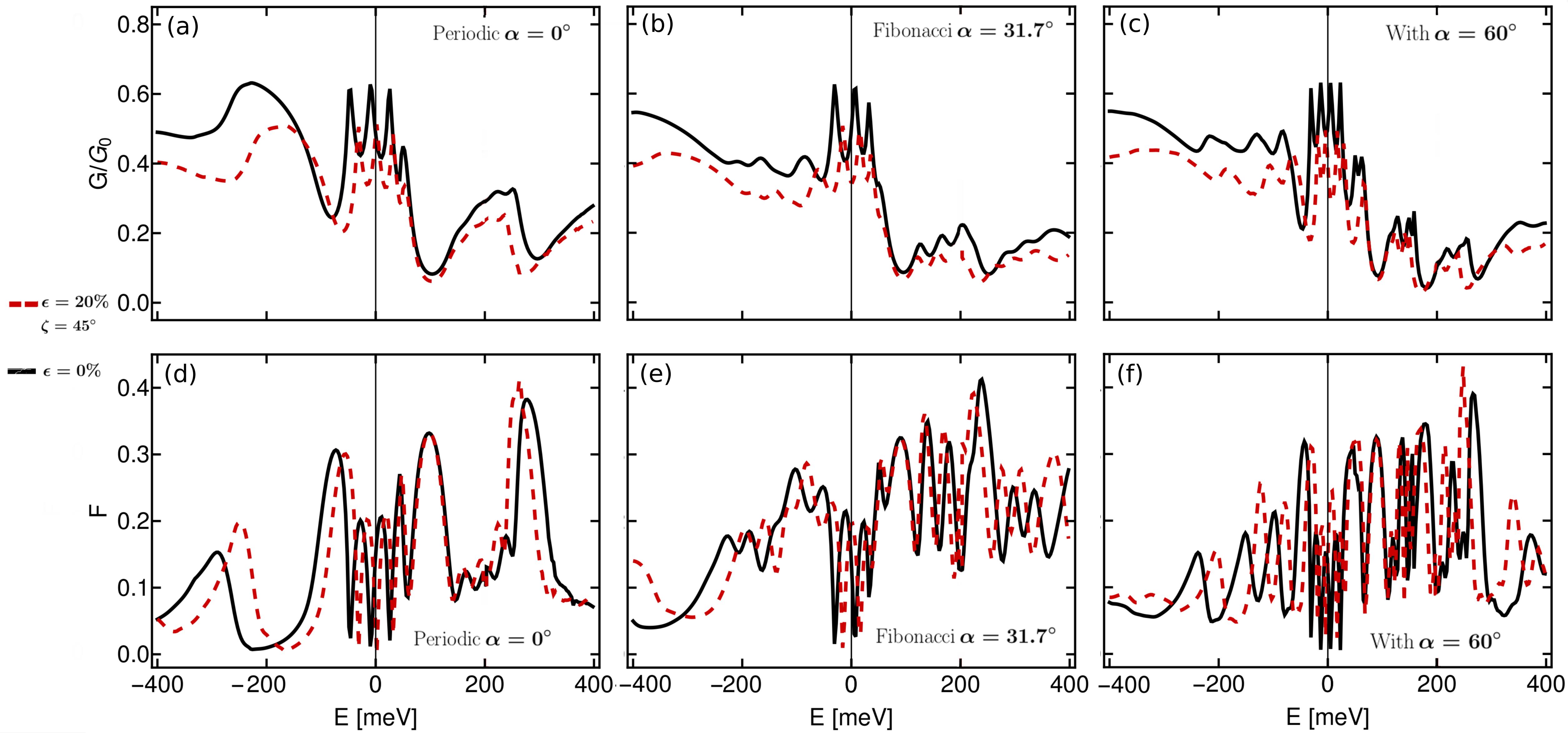}
	\caption{(a)-(c) Conductance $G/G_{0}$ and (d)-(f) Fano factor as function of the Fermi level $E$ for 10 barriers and different values of the parameter $\alpha$. Black and red curves correspond to the cases of pristine and strained graphene, respectively.}
	\label{fig:fig4}
\end{figure}

\section{Results and discussion}\label{sec3}
We discuss the effects of tensile deformations ($\epsilon>0$), applied out the zigzag and armchair directions, on the resonant tunneling of electrons in a finite layer with potential barriers of constant height $V_{0}$ and a periodic order that depends on the parameter $\alpha$ in Eq.~(\ref{sequence}).

The transmission probability $T$, 
as a function of incidence angle $\theta$ and different values of $\alpha$, 
is almost perfect in the whole range $-30^\circ<\theta<0^\circ$, as shown in Fig.~\ref{fig:fig2}(a). Outside this range, $T$ shows smooth oscillations whose amplitudes reach the value of unity for specific values of $\theta$ due to the Fabry-Pérot resonances. In grazing incidence, the transmission probability goes to zero rapidly, delimiting propagation and evanescent modes. Increasing $\alpha$, the high transmission is narrow and decreases abruptly outside the interval $-50^\circ<\theta<20^\circ$. For Fibonacci sequence at $\alpha=31.7^\circ$, the number of resonances is greater than other values of $\alpha$. In Fig.~\ref{fig:fig2}(b), we observe that narrow angular zones appear by increasing the number $N$ of barriers. Such zones resemble pass band effect, where electrons in specific incidence angles can pass perfectly by resonant tunneling. 

Using the Fibonacci sequence for the electrostatic potential, we analyze the effect of tensile strain at the tension angle $\zeta = 45^\circ$, as seen in Fig. \ref{fig:fig2}(c). Starting as reference the case $\epsilon=0\%$, where $T$ is symmetric with respect to the normal incidence, the transmission probability is modified as $\epsilon$ increases. Klein tunneling is shifted to negative incidence angles as well as their resonances. Nevertheless, for positive angles, the other set of resonances remain almost constant. This attributable mainly to the anisotropy of the dispersion relation caused by the strain. 

To complete the effect of strain on graphene with a Fibonacci sequence of barriers, we set $\epsilon=20\%$ and compare the transmission changing the tension angle $\zeta$ in Fig.~\ref{fig:fig2}(d). We can observe how transmission starts being symmetric with respect to $\theta=0^\circ$, even if the graphene layer is uniaxially strained along the zigzag direction, and then the peaks shift for $\theta<0^\circ$ as $\zeta$ increases. Moreover, the resonances are closer for negative incidence angles. Contrary to the case of positive angles, where resonances are separated from each other. It is worth noting that the angular range, where the transmission is almost perfect, become wider when $\zeta=15^\circ$ and $\zeta=30^\circ$.

In Fig.~\ref{fig:fig3}, we show the transmission as a function of the electron incidence energy $E$ and the incidence angle $\theta$. 
The angular region where $T$ is close to unity shifts to negative angles. In contrast with the Klein tunneling in graphene that occurs for normal incidence~\cite{Katsnelson2006}, the strained case has the perfect transmission at $\theta_{\rm KT}\approx-20.92^\circ$. This perfect transmission is not resonant, as shown in Figs. \ref{fig:fig3}(a)-(d), where the aperiodic potential is changed. This is a consequence of the conservation of the pseudo-spin $\phi(\vec{k}^t_0)=\phi(\vec{k}^n_t)$, which is also independent on $E$. The phenomenon, known as anomalous Klein tunneling, is controlled by the strain uniquely \cite{yonatan18}. On the other hand, when the sequence of barriers is periodic ($\alpha=0^\circ$ in Eq.~(\ref{sequence})) and for a fixed incidence angle $\theta$, alternating zones of minimum and maximum value in the transmission coefficient are clearly observed, as the incidence energy increases. In addition, we observe that the transmission coefficient takes on high values for larger angles of incidence when the energy $E$ is less than $V_{0}$, due to the Fabry-Pérot interference. Now, as the parameter $\alpha$ varies until obtaining a Fibonacci sequence for the barriers along the $x$ axis, see Fig.~\ref{fig:fig3}(a)-(c), the regions where resonances occur become more defined for $E>V_{0}$, which implies that their width gets narrow. In contrast, for $E<V_{0}$, there is a wide range of incidence angle values for which transmission is maximum. When $\alpha=45^\circ$, the system is periodic again and the transmission shows a wide angular region for constructive interference. Energy minigaps ($T=0$) emerge for electrons impinging obliquely the barriers with $E < V_{0}$, as a consequence of the destructive interference. The number of energy minigaps continue growing as $\alpha$ increases, as shown in Fig.~\ref{fig:fig3}(d)-(f). This behavior in the transmission is expected because the aperiodicity tends to favor destructive intereference phenomena. With $\alpha=75^\circ$ and $E<V_{0}$, energy minigaps and resonances have more contrast due to that both features follow the aperiodicity of the system.

We further study the conductance per unit length given by
\begin{align}
    G&=\frac{e^2}{\hbar}\int T {\rm d}k_{y}=G_{0}\int_{-\pi/2}^{\pi/2}T\,\frac{\sec^2\theta}{f(\theta)}{\rm d}\theta,
\end{align}
where
\begin{equation}
    f(\theta)=\left(1+\frac{(\tan\theta - \tan\theta_{\rm KT})^2}{\frac{v^2_y}{v^2_x}-\tan^2\theta_{\rm KT}}\right)^{3/2}, \quad
    G_{0}=\frac{e^{2}\vert E\vert}{\pi v_x\hbar^2\left(\frac{v^2_y}{v^2_x}-\tan^2\theta_{\rm KT}\right)},
\end{equation}
as well as the Fano factor,
\begin{equation}
    F=\frac{\int_{-\pi/2}^{\pi/2}T(1-T)\,\frac{\sec^2\theta}{f(\theta)}{\rm d}\theta}{\int_{-\pi/2}^{\pi/2}T\,\frac{\sec^2\theta}{f(\theta)}{\rm d}\theta}.
\end{equation}
Both functions are shown in Fig.~\ref{fig:fig4}. Red (black) curves correspond to the conductance and Fano factor with (without) strain. Deformations cause a slight relative separation among peaks observing oscillations in the conductance and the Fano factor around at the value $E=0$ meV. A notable feature in the conductance transiting from periodic to aperiodic systems is the reduction of the maximum conductance in the $pp'p$ regime $(E < 0)$, as shown in Fig. \ref{fig:fig4}(a)-(c). This can be understood by the emergence of minigaps in the transmission for aperiodic electrostatic barriers, as noted in Fig. \ref{fig:fig3}(a) and (f) for the $pp'p$ regime. On the other hand the Fano factor, understood as a kind of noise-to-signal ratio \cite{GLATTLI2004401,FPennini_2010}, allows us to indicate that electron transmission discussed here shows sub-Poissonian current fluctuations since $F<1$. Moreover, in  
Fig. \ref{fig:fig4}(d)-(e), we observe that the Fano factor presents an inverse behavior to the conductance.

\section{Conclusion}\label{sec4}

In summary, these results evidence that the interplay of strain-engineering and one-dimensional aperiodic potential modify drastically the transport properties in graphene, see Figures~\ref{fig:fig2}-\ref{fig:fig4}. More precisely, the tension angle serves to control the emergence of anomalous Klein tunneling and the parameter $\alpha$ tailors the aperiodicity of the electrostatic potential, which favors the emergence of minigaps in the transmission. Recently, quasiperiodic patterns in graphene and 2D materials have been a topic of growing interest, whose development may lead to the development of novel nanodevices based on anisotropic 2D materials~\cite{Naumis2019,GARCIACERVANTES2015}. Moreover, the extended transfer matrix method exposed here becomes useful to study the combined effects of anisotropy and quasiperiodicity in the conductance and the Fano factor. We think that there is a fruitful research area to apply both strain engineering and quasiperiodic arrangements using other systems, for instance in magnetic graphene superlattices~\cite{BEZERRA2020}. 

\section*{Acknowledgments}
EDB and AR acknowledge valuable discussion with JRF Lima, as well as financial support from CONACYT Project FORDECYT-PRONACES/61533/2020. YBO acknowledges financial support from UNAM-PAPIIT research grant IA106223. This work was also supported by SIP-IPN grant 20220025.

\section*{Data Availability Statement}
The data that support the findings of this study are available from the corresponding author upon reasonable request.

\bibliographystyle{ieeetr}
\bibliography{biblio}

\end{document}